\documentclass{INTERSPEECH2023}
\usepackage{amsmath, graphicx, amssymb, amsfonts, multirow, makecell, array, color, colortbl}
\definecolor{Gray}{gray}{0.9}
\interspeechcameraready
\title{AdaMS: Deep Metric Learning with Adaptive Margin and Adaptive Scale for Acoustic Word Discrimination}
\name{Myunghun Jung, Hoirin Kim}
\address{School of Electrical Engineering, KAIST, Daejeon, Republic of Korea}
\email{kss2517@kaist.ac.kr, hoirkim@kaist.ac.kr}
\begin{document}
\maketitle
%
\begin{abstract}
Many recent loss functions in deep metric learning are expressed with logarithmic and exponential forms, and they involve margin and scale as essential hyper-parameters.
Since each data class has an intrinsic characteristic, several previous works have tried to learn embedding space close to the real distribution by introducing adaptive margins.
However, there was no work on adaptive scales at all.
We argue that both margin and scale should be adaptively adjustable during the training.
In this paper, we propose a method called Adaptive Margin and Scale (AdaMS), where hyper-parameters of margin and scale are replaced with learnable parameters of adaptive margins and adaptive scales for each class.
Our method is evaluated on Wall Street Journal dataset, and we achieve outperforming results for word discrimination tasks.
\end{abstract}
%
\section{Introduction}
\label{sec:introduction}
A successful approach to discriminating two spoken words based on phonetic similarity is to measure the distance between their neural representations, called \textit{acoustic word embeddings} (AWEs) \cite{kamper2016deep, settle2016discriminative}.
Learning AWEs is subsumed into deep metric learning (DML), and the recent progress via introducing \textit{acoustically grounded word embeddings} (AGWEs) \cite{he2017multi, settle2019acoustically, jung2019additional, hu2020multilingual}, which encode phonetic contents of text inputs, is also related to the great success of proxy-based DML \cite{movshovitz2017no, kim2020proxy, jung2022asymmetric}.
Thus, finding the optimal DML or proxy-based DML method can lead to further improvement on acoustic word discrimination task.

In recent DML works that have focused on loss functions while leaving the backbone network the same, the loss functions are generally expressed as the sum of two terms making anchor-positives closer and anchor-negatives farther.
For each term, logarithmic and exponential functions such as softmax, softplus, and log-sum-exp have widely been utilized to handle complicated relations within a batch \cite{movshovitz2017no, kim2020proxy, jung2022asymmetric, wang2019multi, yi2014deep}.
They involve margin and scale, where the margin determines boundaries on embedding space, and the scale controls the intensity of punishment for violations.
These hyper-parameters are uniformly tuned under the assumption that all classes have identical shapes of distributions.
However, the consequently learned embedding space cannot describe the real distribution perfectly since each class has an intrinsic characteristic.

To address the problem, several methods have proposed introducing an adaptive margin.
In \cite{ge2018deep}, a margin for Triplet loss varies with the average distance between positive samples in an anchor class.
In \cite{li2020boosting}, an auxiliary network generates margins for classification loss from pre-trained semantic embeddings of anchor and negative classes.
Since these kinds of methods with margins dependent on certain quantities are deficient in generalization capability, an adaptive margin is defined for each class as a learnable parameter that is jointly optimized with the network \cite{wang2022learning, wu2017sampling, li2020symmetric, liu2019adaptiveface}.
In \cite{wu2017sampling, li2020symmetric, liu2019adaptiveface}, regularization is also employed to induce larger margins as they are preferred for higher discriminability.

While research on the adaptive margin is active as such, an adaptive scale is not considered at all, even though the scale is also the essential component.
Therefore, in this paper, we try to apply the adaptive margin and adaptive scale together.
Specifically, we propose a new but straightforward method called Adaptive Margin and Scale (AdaMS) to give flexibility in training dynamics of DML loss functions by replacing hyper-parameters of margin and scale with learnable parameters of adaptive margins and adaptive scales, respectively.
Then we provide a gradient-based analysis of how the margins and scales change adaptively in affecting the training process simultaneously, which is described with examples.
In addition, following the argument of \cite{ustinova2016learning, zhang2019p2sgrad} that DML loss functions are sensitive to hyper-parameters, we impose non-linear constraints on the intervals where the adaptive margins and adaptive scales can vary.

Our AdaMS method is applied to the current state-of-the-art proxy-based DML loss function, and then the learned embedding space is evaluated on Wall Street Journal (WSJ) dataset for word discrimination tasks.
We demonstrate the effectiveness of the proposed approach by achieving meaningful results outperforming all other methods.
%
\section{Proposed method}
\label{sec:proposed}
In this section, we first review Asymmetric-Proxy (AsyP) loss \cite{jung2022asymmetric}, which reported highly improved results, as the baseline to which AdaMS is applied.
We then describe our AdaMS method and analyze the behaviors of adaptive margins and adaptive scales separately based on their gradients for optimization and an illustration.
\begin{figure*}[ht!]
\begin{minipage}{0.48\linewidth}
\centering
\centerline{\includegraphics[width = 1.0\linewidth]{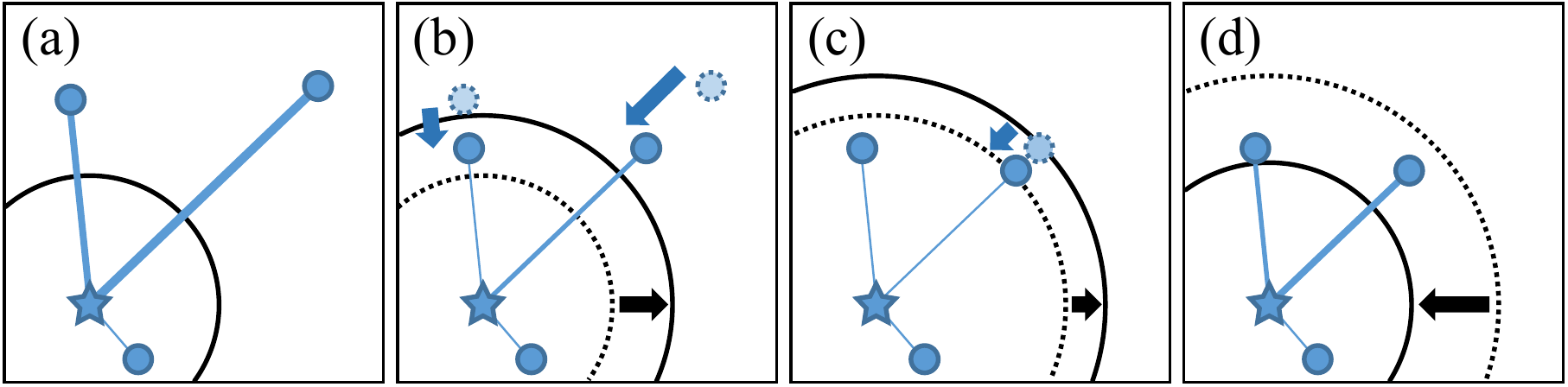}}
\end{minipage}
\hfill
\begin{minipage}{0.48\linewidth}
\centering
\centerline{\includegraphics[width = 1.0\linewidth]{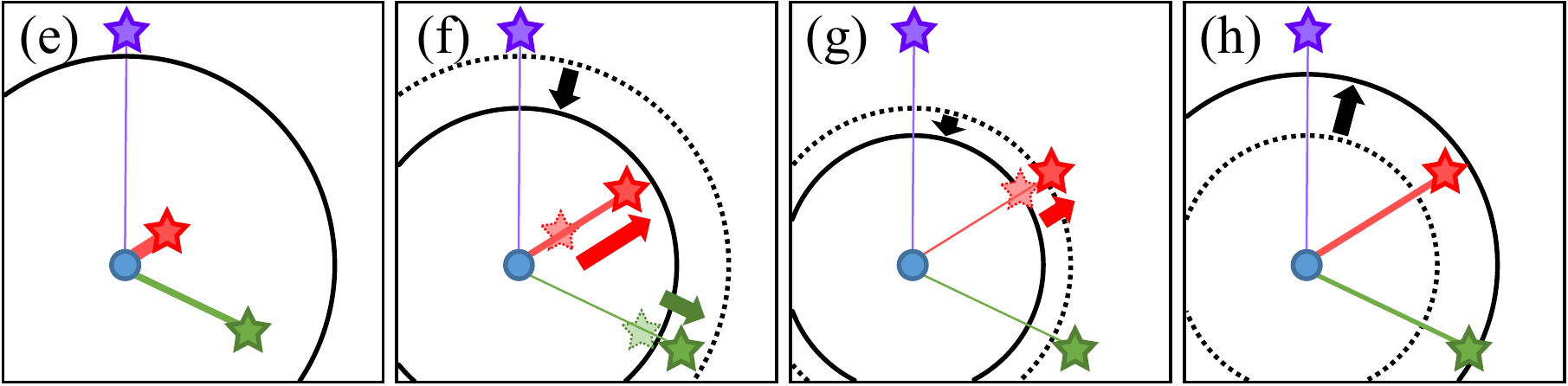}}
\end{minipage}
\caption{Illustration of adaptive margins for positives (a)-(d) and negatives (e)-(h). Small circles are AWEs, stars are AGWEs, and their colors represent distinct classes. The hardnesses are expressed by the thickness of edges. The black solid line indicates the boundary determined by the adaptive margin. All dashes show their previous states, and the movements are represented by arrows.}
\label{fig:1}
\end{figure*}
%
\subsection{Review of Asymmetric-Proxy loss}
\label{ssec:review}
Let $\{ (\mathbf{x}_i, \mathbf{t}_i, c_i) | i = 1, 2, \cdots, N \}$ be a batch of $N$ data tuples, where $\mathbf{x}_i$ is the AWE of the $i$-th speech segment, $\mathbf{t}_i$ is the AGWE of the text label, and $c_i$ is the word class index.
Following the general formulation for proxy-based DML in \cite{jung2022asymmetric}, loss functions are given as the sum of the anchor-positive term $\mathcal{L}^{\mathcal{P}}_i$ and anchor-negative term $\mathcal{L}^{\mathcal{N}}_i$ for the $i$-th anchor:
\begin{equation}
\mathcal{L} = \frac{1}{N} \sum\limits^{N}_{i = 1} \Big( \mathcal{L}_i^{\mathcal{P}} + \mathcal{L}_i^{\mathcal{N}} \Big).
\label{eq:1}
\end{equation}

AsyP loss \cite{jung2022asymmetric} uses two different functions for $\mathcal{L}^{\mathcal{P}}$ and $\mathcal{L}^{\mathcal{N}}$ to show that the optimal function making anchor and positives closer has to be considered separately from one for anchor and negatives to be farther.
Specifically, AsyP loss takes the positive term from Multi-Similarity (MS) loss \cite{wang2019multi} and the negative term from Binomial Deviance (BD) loss \cite{yi2014deep} with modification in computing similarities by introducing proxies, and then combines them.
Each term is given as:
\begin{equation}
\mathcal{L}^{\mathcal{P}}_i = \frac{1}{\alpha} \log \Bigl( 1 + \sum\limits_{j \in \mathcal{P}_i} e^{\alpha (\lambda - S(\mathbf{t}_i, \mathbf{x}_j))} \Bigr),
\label{eq:2}
\end{equation}
\begin{equation}
\mathcal{L}^{\mathcal{N}}_i = \frac{1}{|\mathcal{N}_i|} \sum\limits_{k \in \mathcal{N}_i} \log \left( 1 + e^{\beta (S(\mathbf{x}_i, \mathbf{t}_k) - \lambda)} \right),
\label{eq:3}
\end{equation}
where $\mathcal{P}_i = \left\{ j | c_j = c_i \right\}$, $\mathcal{N}_i = \left\{ k | c_k \neq c_i \right\}$, and $S(\cdot, \cdot)$ denotes the cosine similarity.
$\lambda$ is a fixed margin, $\alpha > 0$ is a fixed scale for the anchor-positive term, and $\beta > 0$ is a fixed scale for the anchor-negative term.
Notice that the proxies $\mathbf{t}$ are utilized as anchors in $\mathcal{L}^{\mathcal{P}}$ and as negatives in $\mathcal{L}^{\mathcal{N}}$.
The gradient of AsyP loss with respect to $S$ is then:\
\begin{equation}
\frac{\partial \mathcal{L}^{\mathcal{P}}_i}{\partial S(\mathbf{t}_i, \mathbf{x}_j)} = \frac{- e^{\alpha (\lambda - S(\mathbf{t}_i, \mathbf{x}_j))}}{1 + \sum\limits_{j' \in \mathcal{P}_i} e^{\alpha (\lambda - S(\mathbf{t}_i, \mathbf{x}_{j'}))}},
\label{eq:4}
\end{equation}
\begin{equation}
\frac{\partial \mathcal{L}^{\mathcal{N}}_i}{\partial S(\mathbf{x}_i, \mathbf{t}_k)} = \frac{\beta}{|\mathcal{N}_i|} \frac{e^{\beta (S(\mathbf{x}_i, \mathbf{t}_k) - \lambda)}}{1 + e^{\beta (S(\mathbf{x}_i, \mathbf{t}_k) - \lambda)}}.
\label{eq:5}
\end{equation}

As described in \cite{kim2020proxy, jung2022asymmetric, wang2019multi}, the magnitude of Equation \ref{eq:4} is determined by the \textit{relative-hardness}.
It means, the gradient is affected by not only the hardness of the similarity to be optimized but also the intra-variance which is expressed as the sum of hardnesses from the positive set.
From the softmax-like form, a hard positive ($S < \lambda$) in low intra-variance class has a highlighted gradient, whereas it has a smoothed gradient with others in high intra-variance class.

For the case of Equation \ref{eq:5}, the magnitude is determined by the \textit{self-hardness} \cite{jung2022asymmetric, wang2019multi}.
As it has the sigmoidal form, the gradient is solely affected by the hardness of the similarity to be optimized.
Also, a hard negative ($S > \lambda$) has a not too much gradient as well as its gradient becomes significantly lower after pushed away.
This property helps the global structure of embedding space to be converged.

However, while AsyP loss takes advantage of MS loss, BD loss, and the concept of proxy-based DML, it cannot account for the effects of hyper-parameters that vary over training because their values are fixed.
\subsection{AdaMS: Adaptive Margin and Scale}
\label{ssec:adams}
Our AdaMS is applied to improve AsyP loss while remaining its inherent advantages.
In order to resolve the limitations, we set the following requirements.
\begin{itemize}
\item Like the scales $\alpha$ and $\beta$, we need two separate margins for positive and negative terms.
\item Margins and scales have to be defined for each class since unified values cannot model various data distributions reflecting their intrinsic characteristics.
\item Margins and scales should be adjustable according to how well the training optimizes local and global structures on embedding space.
\end{itemize}

Our main idea satisfying these requirements is to replace the fixed valued hyper-parameters in Equation \ref{eq:2} and Equation \ref{eq:3} with class-dependent learnable parameters for positive and negative terms separately.
Similarly to \cite{wu2017sampling, li2020symmetric, liu2019adaptiveface}, we also employ regularization for margins.
Resulting loss terms are given as:
\begin{equation}
\mathcal{L}^{\mathcal{P}}_i = \frac{1}{\text{sg}[\alpha_i]} \log \Bigl( 1 + \sum\limits_{j \in \mathcal{P}_i} e^{\alpha_i (\lambda^{\mathcal{P}}_i - S(\mathbf{t}_i, \mathbf{x}_j))} \Bigr) - \omega \lambda^{\mathcal{P}}_i,
\label{eq:6}
\end{equation}
\begin{equation}
\mathcal{L}^{\mathcal{N}}_i = \frac{1}{|\mathcal{N}_i|} \sum\limits_{k \in \mathcal{N}_i} \log \left( 1 + e^{\beta_i (S(\mathbf{x}_i, \mathbf{t}_k) - \lambda^{\mathcal{N}}_i)} \right) + \omega \lambda^{\mathcal{N}}_i,
\label{eq:7}
\end{equation}
where $\text{sg}[\cdot]$ is a stop-gradient operation that prevents the gradient in Equation \ref{eq:4} from changing into an unwanted form, $\omega$ controls the balance of regularizations, and $\lambda^{\mathcal{P}}_i$, $\lambda^{\mathcal{N}}_i$, $\alpha_i$, $\beta_i$ denote the adaptive margins and adaptive scales for class $c_i$.
%
\subsubsection{Behavior of adaptive margin}
\label{sssec:behavior1}
To see how the adaptive margins affect the training, we need to analyze the gradients with respect to $\lambda^{\mathcal{P}}_i$ and $\lambda^{\mathcal{N}}_i$ given as:
\begin{equation}
\frac{\partial \mathcal{L}^{\mathcal{P}}_i}{\partial \lambda^{\mathcal{P}}_i} = \frac{\sum\limits_{j \in \mathcal{P}_i} h^{\mathcal{P}}_{ij}}{1 + \sum\limits_{j \in \mathcal{P}_i} h^{\mathcal{P}}_{ij}} - \omega,
\label{eq:8}
\end{equation}
\begin{equation}
\frac{\partial \mathcal{L}^{\mathcal{N}}_i}{\partial \lambda^{\mathcal{N}}_i} = -\frac{\beta_i}{|\mathcal{N}_i|} \sum\limits_{k \in \mathcal{N}_i} \frac{h^{\mathcal{N}}_{ik}}{1 + h^{\mathcal{N}}_{ik}} + \omega,
\label{eq:9}
\end{equation}
where $h^{\mathcal{P}}_{ij} = e^{\alpha_i (\lambda^{\mathcal{P}}_i - S(\mathbf{t}_i, \mathbf{x}_{j}))}$ and $h^{\mathcal{N}}_{ik} = e^{\beta_i (S(\mathbf{x}_i, \mathbf{t}_k) - \lambda^{\mathcal{N}}_i)}$ denote hardness metrics for class $c_i$ and given similarity.
$h^{\mathcal{P}}_{ij}$ and $h^{\mathcal{N}}_{ik}$ have exponentially large values for hard inputs, otherwise they become significantly smaller, close to 0.

If we ignore $-\omega$ in Equation \ref{eq:8}, then $\frac{\partial \mathcal{L}^{\mathcal{P}}_i}{\partial \lambda^{\mathcal{P}}_i} > 0$ and its magnitude depends on the number of hard positives.
As illustrated in Figure \ref{fig:1}.(a)-(d), high intra-variance causes a large gradient, so $\lambda^{\mathcal{P}}_i$ decreases.
Then with the more relaxed boundary, some less-hard positives, which were not able to meet the margin before, can come into the area.
It leads the model to focus on harder positives.
If most positives are well distributed inside the boundary, then the regularization $-\omega$ becomes dominant and $\lambda^{\mathcal{P}}_i$ increases.
With the more compact area, hard positives may appear again, then the whole process is repeated.

Likewise, if we ignore $\omega$ in Equation \ref{eq:9}, then $\frac{\partial \mathcal{L}^{\mathcal{N}}_i}{\partial \lambda^{\mathcal{N}}_i} < 0$ and its magnitude gradually increases when hard negatives exist.
As illustrated in Figure \ref{fig:1}.(e)-(h), when negatives cross the anchor's boundary, they cause a large gradient, so $\lambda^{\mathcal{N}}_i$ increases.
Then the model narrows the area and pushes the remaining harder negatives outside.
If there is no violations, then the regularization $\omega$ becomes dominant and $\lambda^{\mathcal{N}}_i$ decreases.
By seeking hard negatives again in the broad area, we can achieve higher inter-variance resulting in higher discriminability.
%
\subsubsection{Behavior of adaptive scale}
\label{sssec:behavior2}
Once the adaptive margin determines the boundary, the adaptive scale adjusts how strictly to punish the violations.
Thus their behaviors are highly associated with each other.
It can be figured out from the gradients with respect to $\alpha_i$ and $\beta_i$:
\begin{equation}
\frac{\partial \mathcal{L}^{\mathcal{P}}_i}{\partial \alpha_i} = \frac{1}{\alpha_i} \frac{\sum\limits_{j \in \mathcal{P}_i} (\lambda^{\mathcal{P}}_i - S(\mathbf{t}_i, \mathbf{x}_{j})) h^{\mathcal{P}}_{ij}}{1 + \sum\limits_{j \in \mathcal{P}_i} h^{\mathcal{P}}_{ij}},
\label{eq:10}
\end{equation}
\begin{equation}
\frac{\partial \mathcal{L}^{\mathcal{N}}_i}{\partial \beta_i} = \frac{1}{|\mathcal{N}_i|} \sum\limits_{k \in \mathcal{N}_i} \frac{(S(\mathbf{x}_i, \mathbf{t}_k) - \lambda^{\mathcal{N}}_i) h^{\mathcal{N}}_{ik}}{1 + h^{\mathcal{N}}_{ik}}.
\label{eq:11}
\end{equation}

If some positives violate the margin ($S < \lambda^{\mathcal{P}}_i$) resulting in $\frac{\partial \mathcal{L}^{\mathcal{P}}_i}{\partial \alpha_i} > 0$, we can regard the distribution of the class $c_i$ as less-trained yet or inherently difficult to learn.
In this case, $\alpha_i$ decreases to smooth the level of $h^{\mathcal{P}}$ and prevent the model from focusing on only a few hard positives.
If $\frac{\partial \mathcal{L}^{\mathcal{P}}_i}{\partial \alpha_i} < 0$ that means there are nothing or not many hard positives, then $\alpha_i$ increases to highlight $h^{\mathcal{P}}$ for the outliers and suppress $h^{\mathcal{P}}$ otherwise.
These behaviors resemble the property of the relative-hardness of Equation \ref{eq:4}, so they jointly boost the discriminative training.

The summation of Equation \ref{eq:11} can be divided into the cases of hard negatives ($S > \lambda^{\mathcal{N}}_i$) and the others.
Since $\frac{h^{\mathcal{N}}}{1 + h^{\mathcal{N}}} \simeq 1$ for the hard negatives, the sign of the gradient is determined by most negatives outside the anchor's boundary.
If their hardnesses are not low enough for a reason such that the global structure of embedding space approximated by the set of proxies \cite{movshovitz2017no} is not constructed yet, then $\frac{\partial \mathcal{L}^{\mathcal{N}}_i}{\partial \beta_i} < 0$ and $\beta_i$ increases.
Also it is accompanied by an increase in $\lambda^{\mathcal{N}}_i$.
Thus highly violating proxies are pushed away, which helps establish the global structure.
Otherwise, $\beta_i$ decreases with $\frac{\partial \mathcal{L}^{\mathcal{N}}_i}{\partial \beta_i} > 0$ so that the model can be converged.
%
\subsubsection{Range constraints}
\label{sssec:range}
It is known that the quality of the resulting embedding space is sensitive to the setting of fixed hyper-parameters \cite{ustinova2016learning, zhang2019p2sgrad}.
Similarly, if adaptive margins and adaptive scales are implemented without any constraint on their range, they can lead to unstable training.
So we impose hyperbolic tangent constraints as:
\begin{equation}
\begin{split}
\lambda^{\mathcal{P}}_{i, \text{const}} &= \lambda_0(1 + \tanh (\lambda^{\mathcal{P}}_i))\\
\lambda^{\mathcal{N}}_{i, \text{const}} &= \lambda_0(1 + \tanh (\lambda^{\mathcal{N}}_i))\\
\alpha_{i, \text{const}} &= \alpha_0(1 + \delta_\alpha \tanh (\alpha_i))\\
\beta_{i, \text{const}} &= \beta_0(1 + \delta_\beta \tanh (\beta_i))
\end{split}
\label{eq:12}
\end{equation}
where $\delta_\alpha$ and $\delta_\beta$ control the interval where adaptive scales can vary.
Finally, we apply these constrained adaptive margins and constrained adaptive scales to Equation \ref{eq:6} and Equation \ref{eq:7}.
%
\section{Experiments}
\label{sec:experiments}
In order to demonstrate the effectiveness of AdaMS, we evaluate the learned AWEs and AGWEs on acoustic word discrimination task and cross-view word discrimination task by using Average Precision (AP) metric \cite{he2017multi, settle2019acoustically, jung2019additional, hu2020multilingual, jung2022asymmetric} in percent (\%).

The word-level data is drawn from WSJ dataset \cite{paul1992design} with forced alignment; the train/dev/test sets consist of 639501/ 16839/18274 samples from 13386/3289/3239 unique words.
These splits can be obtained through the WSJ recipe officially distributed by the Kaldi toolkit \cite{povey2011kaldi}.
Then we follow the same feature extraction process as \cite{jung2022asymmetric}.

All methods are implemented with PyTorch \cite{paszke2019pytorch}.
Details that are not mentioned in this paper follow the default setting of PyTorch.
For AWEs and AGWEs, we use two 2-layer BLSTM with 512 units per direction as embedding networks (with dropout of 0.4 only for AWEs) and concatenate the last outputs respectively, so $\mathbf{x}, \mathbf{t} \in \mathbb{R}^{1024}$.
We use Adam optimizer \cite{kingma2015adam} with learning rate of $10^{-4}$ for the embedding networks and $10^{-5}$ for the adaptive margins and adaptive scales.

We set $\lambda = 0.5$, $\alpha = 2$, $\beta = 50$ for the fixed hyper-parameters and for the initial values of the learnable parameters.
In the case of using the range constraints as Equation \ref{eq:12}, we initialize the adaptive margins and adaptive scales to 0 before applying the constraints.
With $\lambda_0 = 0.5$, $\alpha_0 = 2$, $\beta_0 = 50$, $\delta_\alpha = 0.5$, $\delta_\beta = 0.1$, the constrained parameters have the same initial values with non-constrained ones.
All these values are chosen by conducting a grid search with the dev set.
Note that $\omega$ has to differ according to the batch size $N$.
We set $N = 256$, and then $\omega = 0.01$ works well.
%
\subsection{Comparison with other methods}
\label{ssec:comparison}
We compare the performance of our AdaMS with other methods including MV Triplet loss \cite{he2017multi}, MS loss \cite{wang2019multi}, BD loss \cite{yi2014deep}, and the baseline AsyP loss \cite{jung2022asymmetric}.
As originally MS loss and BD loss are not designed for proxy-based DML methods, so we reformulate them as \cite{jung2022asymmetric}.
All experiments are repeated 5 times and we report the mean values with standard deviations.
\begin{table}[h!]
\caption{Word discrimination results on WSJ test set. The subscripts indicate the position of proxies: `P/N' for AGWEs as positives and negatives, `A' for AGWEs as anchors.}
\label{tab:1}
\centering
\begin{tabular}{l|c|c}
\hline
\multirow{2}*{\makecell{Methods}} & \multirow{2}*{\makecell{Acoustic\\AP}} & \multirow{2}*{\makecell{Cross-view\\AP}}\\
 & & \\
\hline
Contrastive & 47.9 (6.86) & -\\
Triplet & 81.2 (1.34) & -\\
MV Triplet & 83.3 (0.72) & 91.0 (0.56)\\
\hline
$\text{Proxy-NCA}_\text{P/N}$ & 86.9 (0.78) & 92.7 (0.52)\\
$\text{Proxy-NCA}_\text{A}$ & 81.5 (0.52) & 89.4 (0.40)\\
\hline
$\text{Proxy-BD}_\text{P/N}$ & 90.5 (0.40) & 95.4 (0.24)\\
$\text{Proxy-BD}_\text{A}$ & 90.8 (0.52) & 95.6 (0.27)\\
\hline
$\text{Proxy-MS}_\text{P/N}$ & 90.8 (0.56) & 96.3 (0.30)\\
$\text{Proxy-MS}_\text{A}$ & 88.1 (0.67) & $\underline{96.4}$ (0.13)\\
\hline
AsyP & $\underline{92.1}$ (0.43) & 96.3 (0.23)\\
\rowcolor{Gray}
AsyP + AdaMS (Ours) & $\mathbf{92.7}$ (0.13) & $\mathbf{96.7}$ (0.12)\\
\hline
\end{tabular}
\end{table}

The comparison results are summarized in Table \ref{tab:1}.
It can be seen that our method outperforms all others on both tasks.
In \cite{jung2022asymmetric}, AsyP loss improved in Acoustic AP compared to other methods.
However, the performance in Cross-view AP was saturated at some level showing little difference with $\text{Proxy-MS}_\text{P/N}$ and $\text{Proxy-MS}_\text{A}$ losses.
When we apply our AdaMS method to AsyP loss, then we can get extra improvements not only in Acoustic AP but also in Cross-view AP.
%
\subsection{Discriminability of unseen words}
\label{ssec:discriminability}
In practical situations of open-vocabulary wake-up word detection or query-by-example spoken term detection, which mainly utilizes AWEs, enrolled or queried word is highly probable to have been unseen at the training.
Therefore, evaluating the discriminability of unseen words is practically meaningful.

In the test set, there are 402 unseen word classes.
To measure the performance in this case, AP for the acoustic word discrimination task is computed by using the samples of unseen words only as the query.
The whole test set is used as the retrieval set as the same.
To simulate the realistic scenario, the cross-view task is not considered since it is plausible that no text labels are provided for unseen words.
\begin{table}[h!]
\caption{Acoustic word discrimination results with unseen word queries on WSJ test set.}
\label{tab:2}
\centering
\begin{tabular}{l|c}
\hline
\multirow{2}*{\makecell{Methods}} & \multirow{2}*{\makecell{Acoustic\\AP}}\\
 & \\
\hline
AsyP & 63.5 (1.72)\\
\rowcolor{Gray}
AsyP + AdaMS (Ours) & $\mathbf{72.8}$ (2.83)\\
\hline
\end{tabular}
\end{table}

The results are given in Table \ref{tab:2}.
The relative improvement about 14.6\% clearly demonstrates the outperforming performance of our AdaMS method.
%
\subsection{Ablation study}
\label{ssec:ablation}
To investigate the importance of each part of the AdaMS, we conduct an ablation study as shown in Table \ref{tab:3}.
\begin{table}[h!]
\caption{Ablation study.}
\label{tab:3}
\centering
\begin{tabular}{l|c|c}
\hline
\multirow{2}*{\makecell{Methods}} & \multirow{2}*{\makecell{Acoustic\\AP}} & \multirow{2}*{\makecell{Cross-view\\AP}}\\
 & & \\
\hline
AsyP (baseline) & 92.1 (0.43) & 96.3 (0.23)\\
\hline
+ adaptive margin & 92.3 (0.37) & 96.4 (0.31)\\
\quad + range constraints & 92.3 (0.22) & 96.3 (0.21)\\
\hline
+ adaptive scale & 91.2 (0.43) & 95.9 (0.22) \\
\quad + range constraints & 91.7 (0.38) & 96.1 (0.21)\\
\hline
+ both & 91.8 (0.21) & 96.3 (0.11)\\
\rowcolor{Gray}
\quad + range constraints (Ours) & 92.7 (0.13) & 96.7 (0.12)\\
\hline
\end{tabular}
\end{table}

First, we apply the adaptive margin and adaptive scale separately.
Similar to \cite{wu2017sampling, li2020symmetric, liu2019adaptiveface}, the adaptive margin shows its own effectiveness.
However, applying the adaptive scale alone leads to a performance degradation.
The reason we can think of is that the decision boundaries are eventually determined by the margins, so if the margins are fixed, then the varying scales rather make the training unstable.
Even if we employ both methods, the result gets better but is still lower than the baseline.
Here, we can understand that the stability of the training is related to the adaptive scale, and this discovery raises the necessity of introducing the range constraints.

When we use the range constraints to the method applying the adaptive margins alone, there is almost no change.
On the other hand, there is an improvement in the adaptive scales with the range constraints.
But it seems there is a limit as well since the adaptive margins are not applied together.
Finally, when we apply the range constraints to the method with the adaptive margins and adaptive scales, we get the best result.

In sum, the class-dependent adaptive margins are effectual by themselves, as \cite{wu2017sampling, li2020symmetric, liu2019adaptiveface}.
To achieve more improvement, we have to apply the adaptive margins and adaptive scales together, but the range constraints are crucial in this case.
%
\subsection{Visualization}
\label{ssec:visualization}
\begin{figure}[t!]
\begin{minipage}{0.48\linewidth}
\centering
\centerline{\includegraphics[width = 1.0\linewidth]{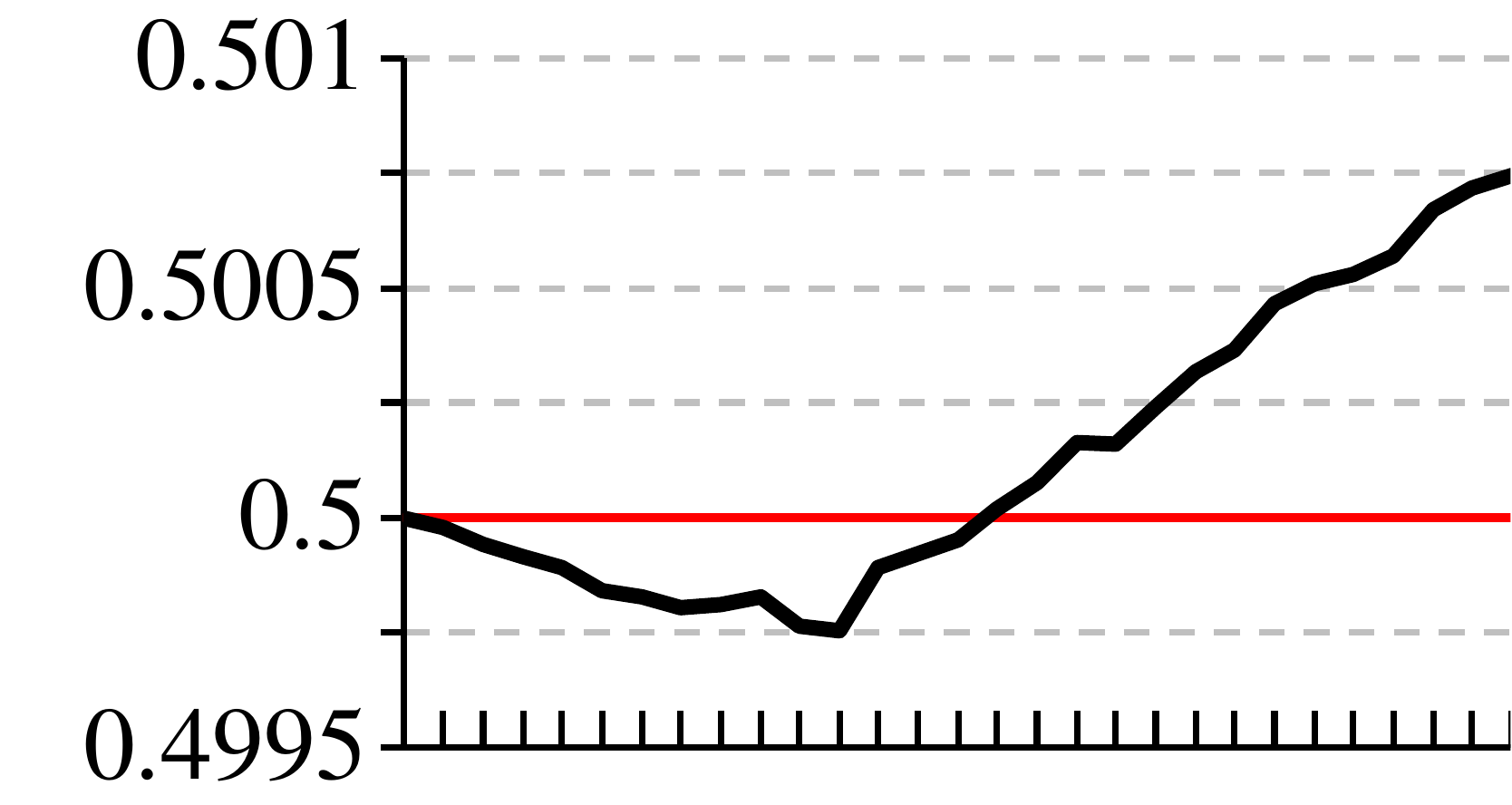}}
\centerline{\quad \quad \quad (a) $\lambda^{\mathcal{P}}_{\texttt{behavior}}$}
\end{minipage}
\hfill
\begin{minipage}{0.48\linewidth}
\centering
\centerline{\includegraphics[width = 1.0\linewidth]{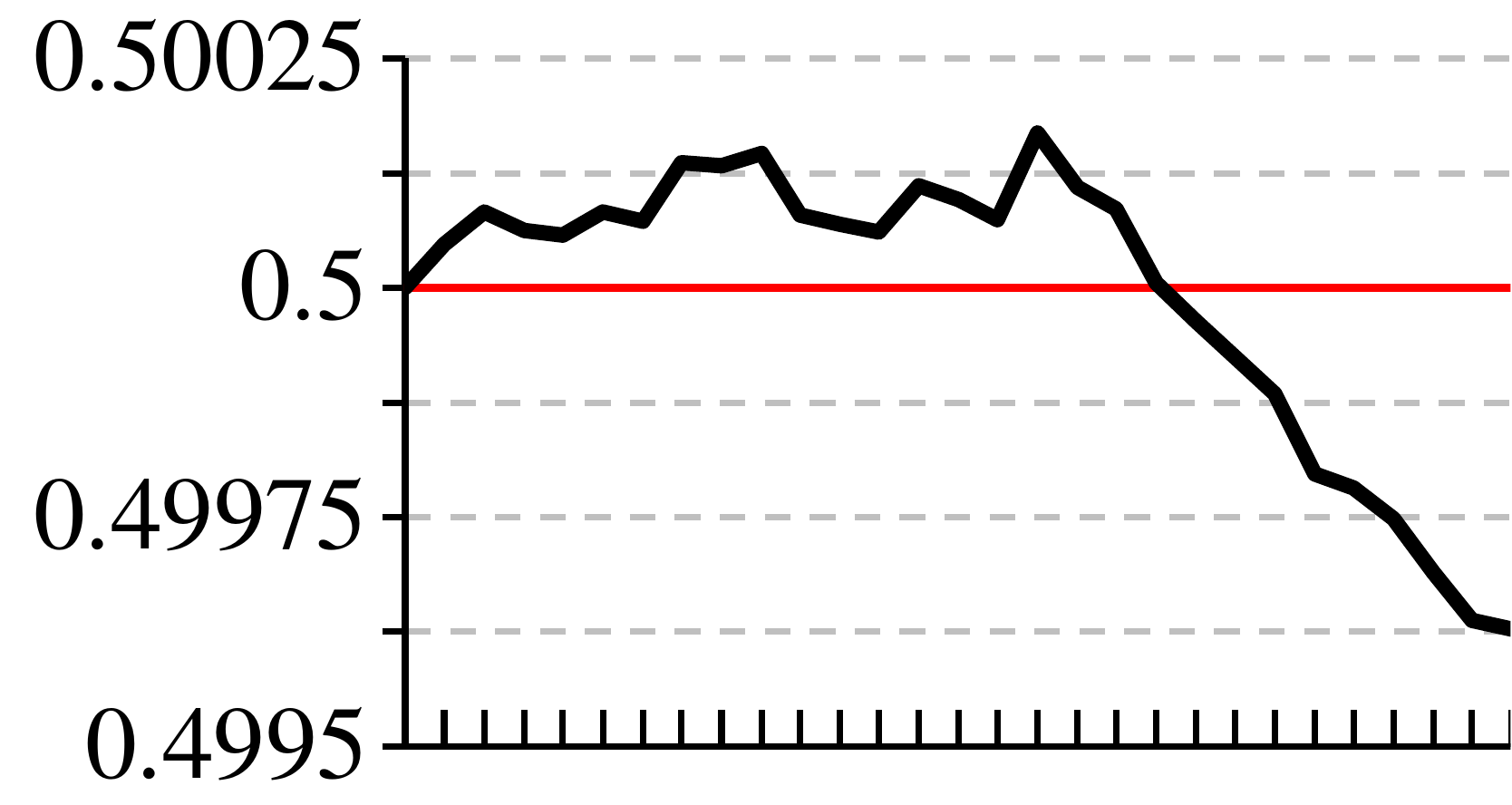}}
\centerline{\quad \quad \quad (b) $\lambda^{\mathcal{N}}_{\texttt{behavior}}$}
\end{minipage}\\
\vspace{0.3cm}\\
\begin{minipage}{0.48\linewidth}
\centering
\centerline{\includegraphics[width = 1.0\linewidth]{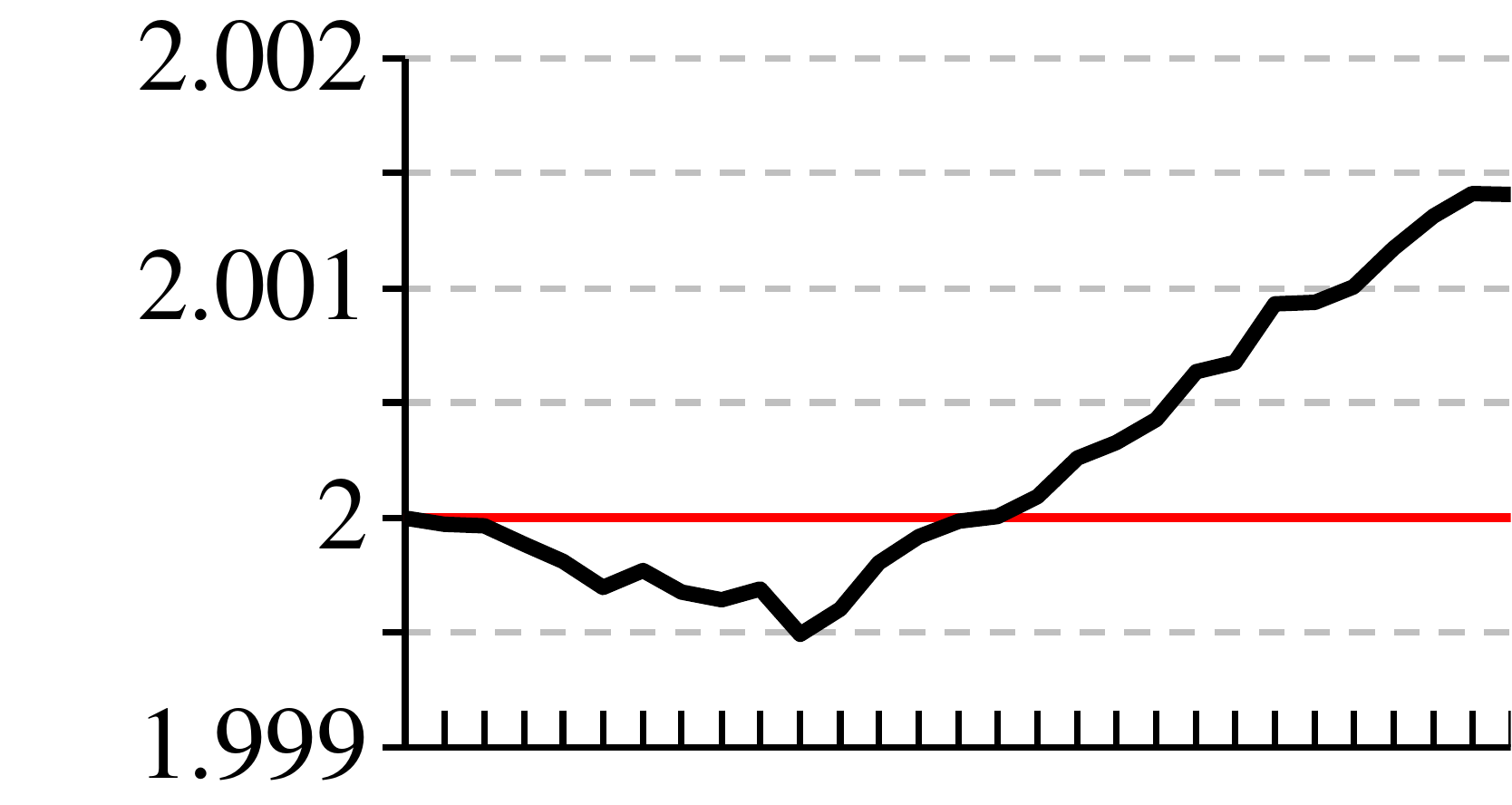}}
\centerline{\quad \quad \quad (c) $\alpha_{\texttt{behavior}}$}
\end{minipage}
\hfill
\begin{minipage}{0.48\linewidth}
\centering
\centerline{\includegraphics[width = 1.0\linewidth]{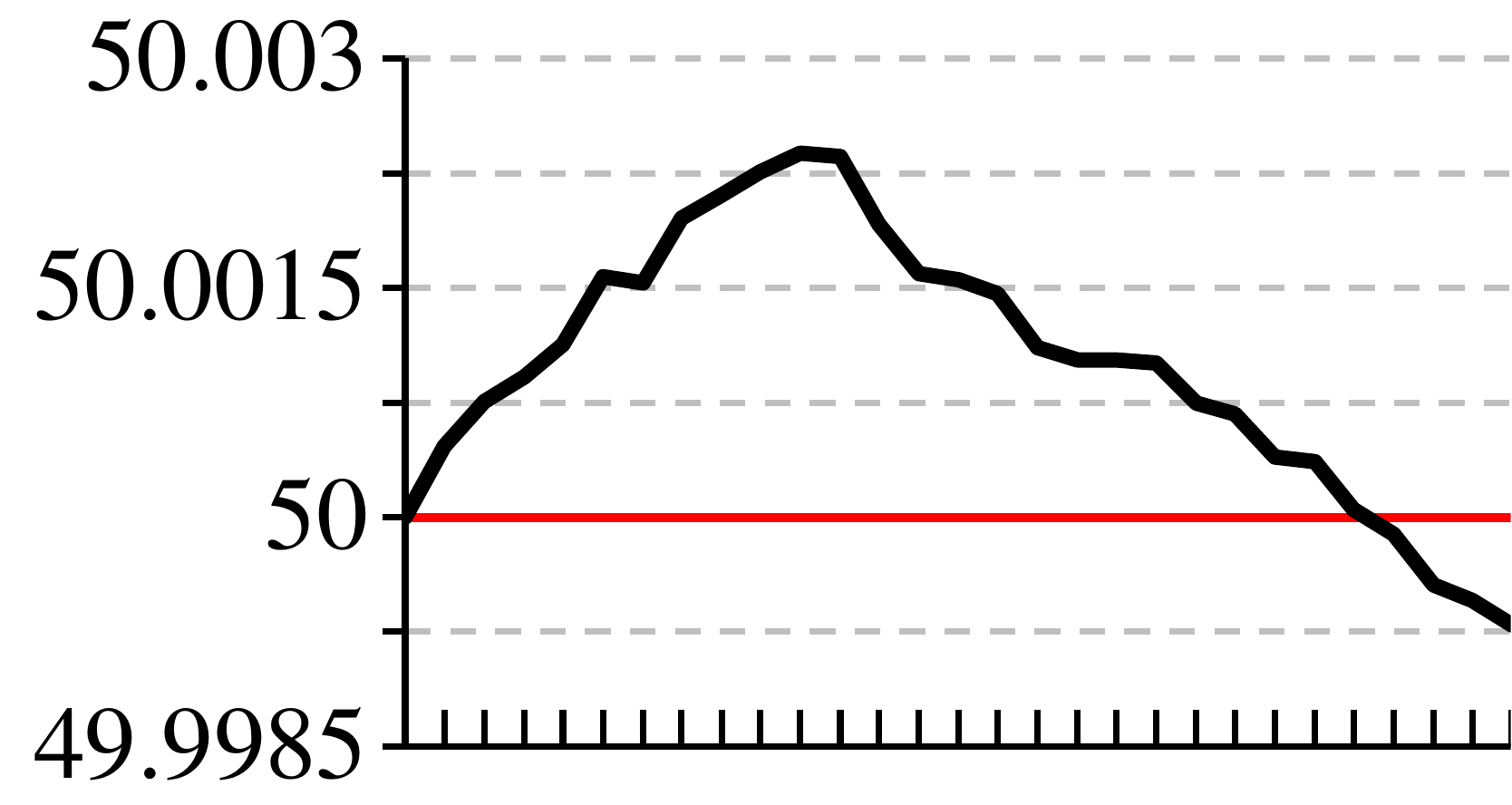}}
\centerline{\quad \quad \quad (d) $\beta_{\texttt{behavior}}$}
\end{minipage}
\caption{Visualization of how the adaptive margins and adaptive scales for the word `\texttt{behavior}' change during the first epoch. Red horizontal lines indicate the initial values.}
\label{fig:2}
\end{figure}
In order to see the actual behavior of the adaptive margins and adaptive scales described in Sec.\ref{ssec:adams}, all values are recorded during the training.
Particularly, we focus only on the first epoch.
Because all model parameters rapidly move to the nearest optima with the steepest gradients at the beginning, we can observe their distinct behavior.

We select the word class `\texttt{behavior}' and visualize its values in Figure \ref{fig:2}.
In this example, as we expected, $\lambda^{\mathcal{P}}$ is highly correlated with $\alpha$, and they decrease until a particular moment and then increase.
Also, though $\lambda^{\mathcal{N}}$ fluctuates a bit, it shows a similar trend with $\beta$, where they increase at first and then decrease.
The fluctuation of $\lambda^{\mathcal{N}}$ can be interpreted that the development of the global structure of the embedding space at the early stage of the training incurs violations and punishing frequently.
%
\section{Conclusion}
\label{sec:conclusion}
In this paper, we have proposed a novel AdaMS method to address the problem of fixed hyper-parameters in DML loss functions.
It involves simple modification by replacing hyper-parameters of margin and scale with range constrained learnable parameters of adaptive margins and adaptive scales.
Since each class can approximate its real distribution closely, we can construct more discriminative embedding space.
The outperforming results on word discrimination tasks demonstrate the effectiveness of our method.
In addition, through the ablation study, we have demonstrated that applying adaptive margins and adaptive scales together is meaningful and that the range constraints need to be considered at the same time.
%
\section{Acknowledgement}
%
This work is/was supported by Samsung Research, Samsung Electronics Co., Ltd.
%
\bibliographystyle{IEEEtran}
\bibliography{mybib}
\end{document}